\documentclass[sigconf]{acmart}
\AtBeginDocument{%
  }

\usepackage{xcolor}
\usepackage{soul} 
\usepackage{makecell}
\usepackage{xspace}
\usepackage{listings}
\usepackage{hyperref}
\usepackage{subcaption}
\usepackage[most]{tcolorbox}
\tcbset{
  colback=gray!5,
  colframe=gray!40!black,
  boxrule=0.5pt,
  arc=2pt,
  left=6pt, right=6pt, top=4pt, bottom=4pt
}
\usepackage{soul}
\setulcolor{blue}
\newcommand{\ulhref}[2]{\href{#1}{\ul{#2}}}

\newcommand{\ts}{{Type\-Script}\xspace}
\newcommand{\js}{{Java\-Script}\xspace}

\newcommand{\code}[1]{\text{\lstinline[basicstyle=\ttfamily\small, language=Java]~#1~}}

\definecolor{light_green}{rgb}{0.3, 0.7, 0.33}
\definecolor{light_red}{rgb}{0.7, 0.32, 0.31}

\definecolor{sh_comment}{rgb}{0.12, 0.38, 0.18}
\definecolor{sh_keyword}{rgb}{0.37, 0.08, 0.25}  
\definecolor{sh_string}{rgb}{0.06, 0.10, 0.98} 
\def\lstsmallmath{\leavevmode\ifmmode \scriptstyle \else  \fi}
\def\lstsmallmathend{\leavevmode\ifmmode  \else  \fi}
\definecolor{KWColor}{rgb}{0.5,0,0.67}
\definecolor{CommentColor}{rgb}{0.15,0.5,0.15}
\definecolor{lightgrey}{rgb}{0.8,0.8,0.8}

\lstdefinelanguage{JavaScript}[]{Java}{
   morekeywords={var,class,object,function, async, await, undefined, let, form, button, div, useState, number, textarea, then} 
} 

\lstdefinestyle{Eclipse}{
  xleftmargin=0pt,
  language = JavaScript,
  basicstyle=\sffamily\footnotesize,
  stringstyle=\color{sh_string},
  keywordstyle = \color{sh_keyword}\bfseries,  
  lineskip=-0.0em,
  commentstyle=\color{sh_comment}\itshape,  
  escapeinside={/*@}{@*/},
  numbersep=5pt,
  captionpos=b,
  xleftmargin=0.4cm, xrightmargin=0.5cm,
   morekeywords={invokestatic,invokeinterface,invokevirtual,invokespecial,then},
}

\hypersetup{
  colorlinks=true,
  linkcolor=blue,
  urlcolor=blue,
  citecolor=blue
}

\newcommand\numOfCategories{11\xspace}
\newcommand\numOfProjects{16\xspace}
\newcommand\numOfBugs{633\xspace}

\copyrightyear{2026}
\acmYear{2026}
\setcopyright{cc}
\setcctype{by}
\acmConference[MSR '26]{23rd International Conference on Mining Software Repositories}{April 13--14, 2026}{Rio de Janeiro, Brazil}
\acmBooktitle{23rd International Conference on Mining Software Repositories (MSR '26), April 13--14, 2026, Rio de Janeiro, Brazil}
\acmPrice{}
\acmDOI{10.1145/3793302.3793379}
\acmISBN{979-8-4007-2474-9/2026/04}

\begin{document}

\title{From Logic to Toolchains:\\An Empirical Study of Bugs in the TypeScript Ecosystem}


\author{TianYi Tang}
\affiliation{%
  \institution{Simon Fraser University}
  \city{Burnaby}
  \state{BC}
  \country{Canada}
}
\email{tta89@sfu.ca}

\author{Saba Alimadadi}
\affiliation{%
  \institution{Simon Fraser University}
  \city{Burnaby}
  \state{BC}
  \country{Canada}
}
\email{saba@sfu.ca}

\author{Nick Sumner}
\affiliation{%
  \institution{Simon Fraser University}
  \city{Burnaby}
  \state{BC}
  \country{Canada}
}
\email{wsumner@sfu.ca}

\renewcommand{\shortauthors}{Tang et al.}

\begin{abstract}


\ts has rapidly become a popular language for modern web development, yet its effect on software faults remains poorly understood.  
This paper presents the first large-scale empirical study of bugs in real-world \ts projects.  
We analyze \numOfBugs bug reports from \numOfProjects popular open-source repositories to construct a taxonomy of fault types, quantify their prevalence, and relate them to project characteristics such as size, domain, and dependency composition.  
Our results reveal a fault landscape dominated not by logic or syntax errors but by tooling and configuration faults, API misuses, and asynchronous error-handling issues.  
We show that these categories correlate strongly with build complexity and dependency heterogeneity, indicating that modern failures often arise at integration and orchestration boundaries rather than within algorithmic logic.  
A longitudinal comparison with \js studies shows that while static typing in \ts has reduced traditional runtime and type errors, it has shifted fragility toward build systems and toolchains.  
These findings offer new insight into how language design and ecosystem evolution reshape the fault profiles of large-scale software systems.

\end{abstract}




\maketitle

\section{Introduction}
\label{sec:introduction}

\ts has redefined modern web development by extending \js with static typing, modularization, and an extensive compiler and tooling ecosystem. It now underpins major frameworks and platforms such as 
Angular\footnote{\url{https://angular.dev/}} and VS~Code.\footnote{\url{https://code.visualstudio.com/}} By integrating static types into a dynamic language, \ts promises stronger correctness guarantees, improved maintainability, and a more predictable developer experience~\cite{Gao_2017,Bogner_2022}. Yet, despite its rapid adoption, little is known about whether these promises have translated into fewer or different kinds of faults in practice.

Earlier empirical studies of \js bugs~\cite{Ocariza_2011, Ocariza_2013, Ocariza_2017} captured an era of dynamically-typed, browser-centric systems with limited build tooling. Since then, the web ecosystem has evolved dramatically: applications now rely on layered toolchains (e.g., linters, transpilers, bundlers), asynchronous runtimes, and dependency-rich architectures. These shifts likely alter how faults emerge, propagate, and are detected in practice. Understanding this evolution is essential for researchers developing analysis and testing tools, and for practitioners maintaining large-scale \ts systems.

Despite \ts\!\!’s significance, no large-scale, systematic study has examined how faults manifest in \ts-based software. Prior research has either focused on bugs in \js applications~\cite{Ocariza_2011, Ocariza_2013, Ocariza_2017, Gyimesi_2020, Wang_2017}, on the role of static typing and its impact on code quality~\cite{Bogner_2022,Gao_2017,emmanni2021role}, or on defects in the \ts compiler itself~\cite{Wang_2025}. However, these works do not characterize the \emph{fault landscape} of modern \ts projects; i.e., how different kinds of bugs occur, how they relate to project-level characteristics, and how they compare to the \js era.

Specifically, we lack evidence about  
(1) the relative frequency and nature of distinct fault categories (e.g., type, logic, asynchronous, configuration);  
(2) the influence of project-level attributes such as size, domain, and dependency complexity on these fault distributions; and  
(3) whether \js-era bug types persist despite static typing and advanced tooling, or whether new categories of faults have emerged that warrant attention.  
This absence of evidence limits our understanding of how \ts and its ecosystem have reshaped real-world software reliability.

To address this gap, we perform a mixed-method empirical study of bugs in modern, open-source \ts projects. Our dataset includes \numOfBugs bug-related artifacts from \numOfProjects actively maintained repositories across diverse domains (e.g., frameworks, CLIs, libraries, UI systems, and AI integrations). We combine  
(1) \textit{manual and semi-automated labeling}, using iterative human analysis and few-shot learning to construct a taxonomy of \ts bug categories;  
(2) \textit{quantitative analysis}, relating bug categories to project-level factors such as size, domain, and dependency complexity; and  
(3) \textit{longitudinal comparison}, mapping our taxonomy to prior \js studies to examine how web application faults have evolved in the presence of static typing and a more complex tooling ecosystem.

Together, these analyses allow us to characterize the dominant fault categories in modern \ts projects, understand how they vary with ecosystem and architectural factors, and assess how today’s fault landscape differs from that observed in earlier \js systems.


Our results reveal several trends:  
(1) Tooling and configuration issues, rather than type or logic faults, dominate many mature projects;  
(2) Asynchrony- and event-handling-related bugs persist across multiple domains despite \ts\!\!’s type system;
(3) Fault distributions vary systematically with project domain and dependency structure; and  
(4) While \ts reduces some categories of runtime errors, it introduces new forms of build-time and integration fragility.  
%

This paper makes the following contributions:
\begin{itemize}
  \item An empirically-derived taxonomy of \ts bugs capturing both traditional and ecosystem-specific faults;
  \item A quantitative correlation analysis linking bug categories to project characteristics, dependency complexity, and domain;
  \item A longitudinal comparison with prior \js studies, showing how static typing and modern toolchains reshape the web fault landscape; and
  \item A curated open-source benchmark dataset and open analysis pipeline to support reproducible future research on \ts reliability. All scripts and data are publicly available at \ulhref{https://github.com/SEatSFU/tsbugs}{https://github.com/seatsfu/tsbugs}.
\end{itemize}

These contributions provide the first evidence-based characterization of how faults arise and evolve in the \ts ecosystem. They establish a foundation for understanding the reliability challenges of typed web development and for guiding future tools for testing, debugging, and analysis.

\section{Study Design and Methodology}
\label{sec:methodology}

The goal of this study is to empirically characterize the fault landscape of modern \ts projects and explain how it differs from that of earlier \js systems.  
We aim to
(1)\,identify recurring fault types and their prevalence,
(2)\,analyze how these faults relate to intrinsic and ecosystem-level factors like scale, domain, and dependency diversity, and
(3)\,examine how the introduction of static typing and modern build pipelines has shifted the distribution of web application faults.
Together, these analyses describe and explain how web-development bugs have evolved \emph{from logic to toolchains}.


We adopted a mixed-method design combining qualitative analysis, quantitative correlation, and historical comparison.  
The study proceeded in four stages:  
(1)~curation of a diverse corpus of open-source \ts projects and their bug-related commits;  
(2)~manual and semi-automated construction of an empirically grounded bug taxonomy;  
(3)~quantitative analyses linking bug types to project-- and ecosystem-level characteristics; and  
(4)~a comparative assessment of how the web fault landscape has evolved from \js to \ts.  
This pipeline, spanning data collection, taxonomy development, correlation, and historical comparison, enabled both descriptive and explanatory insights into the causes of \ts faults.

\begin{table}[t!]
\footnotesize
\centering
\caption{Overview of Analyzed TypeScript Projects}
\label{tab:benchmarks}
\renewcommand{\arraystretch}{1.1}

\resizebox{\columnwidth}{!}{%
\begin{tabular}{l l r r r r r}
\toprule
\textbf{Project} & \textbf{Domain} & \textbf{LOC} & \textbf{Stars} & \textbf{Commits} & \textbf{Issues} & \textbf{Age (Y)} \\
\midrule
\href{https://github.com/refly-ai/refly }{refly} & AI Integration & 237099 & 4718 & 7138 & 214 & 1.7 \\
\href{https://github.com/run-llama/LlamaIndexTS}{LlamaIndexTS} & AI Integration & 59819 & 2907 & 2390 & 452 & 2.4 \\
\href{https://github.com/Kong/insomnia }{insomnia} & API/service integration & 130711 & 37288 & 5745 & 3905 & 9.5 \\
\href{https://github.com/xyflow/xyflow}{xyflow} & API/service integration & 27224 & 32241 & 5843 & 2295 & 6.3 \\
\href{https://github.com/react-native-community/cli}{cli} & API/service provider & 19011 & 2750 & 2686 & 1160 & 7.1 \\
\href{https://github.com/ueberdosis/hocuspocus}{hocuspocus} & API/service provider & 9690 & 1868 & 1811 & 372 & 4.9 \\
\href{https://github.com/n8n-io/n8n}{n8n} & App. logic \& workflows & 1186492 & 147008 & 15843 & 5993 & 6.3 \\
\href{https://github.com/statelyai/xstate}{xstate} & App. logic \& workflows & 70982 & 28790 & 6841 & 1397 & 10.1 \\
\href{https://github.com/antiwork/shortest}{shortest} & Automated testing \& QA & 15412 & 5328 & 747 & 120 & 1.1 \\
\href{https://github.com/testing-library/user-event}{user-event} & Automated testing \& QA & 11210 & 2286 & 825 & 551 & 7.1 \\
\href{https://github.com/type-challenges/type-challenges}{type-challenges} & Education & 5162 & 46645 & 1016 & 36524 & 5.2 \\
\href{https://github.com/nestjs/nest}{nest} & Library/Framework & 82983 & 72973 & 19066 & 5689 & 8.7 \\
\href{https://github.com/intlify/vue-i18n }{vue-i18n} & Library/Framework & 29457 & 2559 & 3283 & 630 & 4.8 \\
\href{https://github.com/chakra-ui/chakra-ui}{chakra-ui} & User interface & 103193 & 39789 & 11168 & 4189 & 6.2 \\
\href{https://github.com/fabricjs/fabric.js}{fabric.js} & User interface & 67414 & 30467 & 5256 & 6189 & 15.4 \\
\href{https://github.com/sadmann7/shadcn-table}{shadcn-table} & User interface & 9511 & 5466 & 1127 & 74 & 2.3 \\
\midrule
Mean & -- & 129086 & 28220 & 5674 & 4360 & 6.2 \\
Total & -- & 2065370 & -- & 90785 & 69754 & -- \\
\bottomrule
\end{tabular}%
}
\end{table}

\subsection{Research Questions}

Our study is guided by our research questions that together examine the nature, context, and evolution of faults in modern \ts systems:

\noindent\textbf{RQ1:} What are the common categories and prevalence of bugs in modern \ts projects?\\
\noindent\textbf{RQ2:} How do these bug categories relate to project-level characteristics and ecosystem complexity?\\
\hspace*{1em}\textbf{RQ2a:} How are bug categories correlated with intrinsic project attributes such as domain, size, age, activity, and popularity?\\
\hspace*{1em}\textbf{RQ2b:} What relationships exist among different bug categories?\\
\hspace*{1em}\textbf{RQ2c:} How do ecosystem-level factors, such as build tooling and dependency diversity, correlate with specific fault types?\\
\noindent\textbf{RQ3:} How have web application faults evolved from \js to \ts, and what role does static typing play in this evolution?

These questions collectively guide the data collection, taxonomy construction, quantitative analysis, and historical comparison described in the remainder of this section.

\subsection{Dataset Construction and Project Selection}

We mined \numOfBugs bug-related commits, issues, and pull requests from \numOfProjects actively maintained GitHub repositories written primarily in \ts (\textgreater80\% TypeScript files).  
Projects were selected to maximize diversity: each had at least 1{,}000 stars, publicly-visible issues and pull requests, and recent maintenance activity, defined as at least 50 commits to the main branch in the 12 months preceding data collection.
To ensure ecosystem coverage and reduce front-end bias, repositories were selected using stratified sampling across domains, including UI frameworks, developer tools, backend services, automation frameworks, education, and AI integration. We selected two projects per domain (three for UI due to its diversity, and one for education), yielding 16 repositories in total.
For each project, we collected metadata including LOC, stars, commits, pull requests, and project age, which was stored in structured form and later merged with labeled bug data for correlation analyses. \autoref{tab:benchmarks} summarizes the analyzed projects, including their domains, sizes, and activity metrics.

\begin{table*}[t!]
\centering
\footnotesize
\caption{Empirically-derived taxonomy of bug categories in \ts projects.}
\label{tab:taxonomy}
\renewcommand{\arraystretch}{1.1}
\setlength{\tabcolsep}{4pt}

\begin{tabular}{p{0.13\linewidth} p{0.47\linewidth} p{0.33\linewidth}}
\toprule

\textbf{Bug Class} & \textbf{Definition} & \textbf{Example} \\

\midrule

\textbf{Async / Event} & Incorrect use of asynchronous constructs, event handlers, or concurrency & Race condition between event handler and async fetch \\
\textbf{Error Handling} & Absent, incorrect, or overly-broad error handling & Missing a \texttt{try/catch} block \\
\textbf{Missing Case} & Failure to handle a valid input, state, or execution path & Reducer omits a case for a new enum value \\
\textbf{Missing Feature} & Functionality described in requirements or implied by design is unimplemented & API endpoint declared but handler not implemented \\
\textbf{Runtime Exception} & Faults causing crashes at runtime & Invoking a method on \texttt{undefined} or \texttt{null} \\
\textbf{Tooling / Config} & Bugs due to build system, compiler, or configuration mismanagement & \texttt{tsconfig.json} excludes files needed for compilation \\
\textbf{Type Error} & Incorrect or overly-permissive type annotations, unsafe casts, or misuse of generics & Assigning a string to a variable typed as \texttt{number}; bypassing checks with `\texttt{as any}` \\
\textbf{Logic Error} & Faulty computation, branching, or iteration unrelated to types or APIs & Using \texttt{>} instead of \texttt{<} in a loop guard \\
\textbf{API Misuse} & Incorrect invocation of internal or third-party APIs & Passing arguments in the wrong order to a library call \\
\textbf{Test Fault} & Bugs in the test code or test infrastructure & Flaky test caused by improper async handling \\
\textbf{UI Bug} & Unexpected or inconsistent visual/interaction behaviour & Button click fails to update the UI in a React component \\

\bottomrule

\end{tabular}
\end{table*}

\paragraph{Bug collection and filtering.}
For each repository, we extracted commit titles, messages, modified files, and linked issues or pull requests.  
Commits were included if their titles or messages matched fix-related keywords (e.g., \texttt{fix}, \texttt{fixed}, \texttt{fixes \#ID}) after tokenization and lemmatization, and excluded using non-functional indicators (e.g., refactoring, formatting, typos). Robot-generated commits and duplicates were removed.
All text was preprocessed by lowercasing, tokenization, and removing punctuation, code snippets, and stop words to prepare artifacts for classification. This process yielded a curated corpus of unique bug-related artifacts across all projects.

\paragraph{Labeling pipeline.}
A subset of reports was manually annotated by two researchers to construct the initial taxonomy and seed the labeling process.
The researchers independently labeled an overlapping subset, with disagreements resolved through discussion until consensus.
Using this data set, we fine-tuned a compact sentence-embedding model (\texttt{TaylorAI/bge-micro-v2}) in a few-shot configuration to label the remaining artifacts.
Low-confidence predictions (typically below 0.6--0.7) were manually inspected, and targeted spot checks were performed on high-confidence predictions across categories.
This iterative human-in-the-loop process refined category boundaries and produced a balanced, scalable, and reproducible labeled dataset of \numOfBugs bug instances used in subsequent analyses.

\subsection{Taxonomy Development}
\label{sec:taxonomy_construction}

We developed a new bug taxonomy tailored to modern \ts projects using a grounded-theory approach.  
While prior \js taxonomies (e.g., \citet{Ocariza_2013}) informed our analysis, they predate \ts and do not capture several fault types that arise from \ts's type system and today’s web tooling ecosystem. Additional categories therefore emerged empirically from our dataset.
Two annotators independently labeled bug reports using \textit{Label Studio},\footnote{\url{https://labelstud.io/}} examining issue descriptions, linked pull requests, and code diffs. Disagreements were resolved through discussion until consensus was reached.  
Following grounded-theory principles, we applied \emph{open coding} to identify recurring fault patterns and \emph{axial coding} to consolidate them into higher-level categories.
The resulting taxonomy consists of eleven bug categories, capturing both \ts-specific faults (e.g., \emph{Type Error}, \emph{Tooling / Configuration}, \emph{Asynchrony / Event Handling}) and more general software faults (e.g., \emph{Logic Error}, \emph{API Misuse}, \emph{Missing Feature}).  
This taxonomy forms the foundation for the analyses in RQ1 and RQ2.

\subsection{Quantitative Analyses}
\label{sec:quantitative_analyses}

To address RQ2, we analyzed relationships between bug categories and project characteristics.  
Each labeled bug was mapped to its repository, and category counts were aggregated per project.  
To avoid over-representing large or highly-active systems, we sampled comparable numbers of bug instances per project rather than exhaustively mining all issues, controlling for scale bias without requiring normalization by KLOC or commit count.

We used \emph{Spearman’s rank correlation} ($\rho$) to assess associations between numeric project attributes (LOC, commits, stars, age) and bug frequencies, and applied \emph{Kruskal--Wallis tests} ($\alpha=0.05$) to evaluate differences across project domains.  
These non-parametric methods were selected because project metrics and per-category bug counts are highly skewed and non-normal, violating the assumptions of Pearson correlation and ANOVA.  
Z-score normalization was used only to standardize medians for visualization of domain-level over- and under-represented categories.
Results are summarized using correlation heatmaps and stacked distributions, enabling us to relate fault patterns to project maturity, architectural structure, and ecosystem complexity.

\subsection{Historical Comparison RQ3}
\label{sec:historical_mapping}

To assess how the web fault landscape has evolved, RQ3 compares the prevalence and nature of bug categories in our \ts taxonomy 
with those reported in prior studies of \js~\cite{Ocariza_2013, Ocariza_2017}.  
We adopt \citet{Ocariza_2017} as the primary baseline due to its comprehensive and widely-cited characterization of bugs.

\paragraph{Category mapping.}
Because the taxonomies differ, we manually established conceptual correspondences between \ts and \js categories.
Each \ts category was aligned with one or more \js categories sharing similar fault semantics (e.g., ``unhandled exceptions'' in \js partially correspond to our \emph{Error Handling} and \emph{Runtime Exception} categories).
Mappings were independently reviewed and refined through consensus to capture both direct and partial overlaps.

\paragraph{Prevalence comparison.}
Since the mappings are not one-to-one, direct quantitative comparison is infeasible.
Instead, we estimate a \emph{maximum corresponding prevalence} for each \ts category by summing the relative frequencies of its mapped \js categories.
For each \ts category \(C_i\), we identify its mapped set \(M_i\) and compute the cumulative proportion of bugs represented by \(M_i\).
This conservative approach highlights categories that have increased, declined, or newly emerged over time, enabling a qualitative assessment of ecosystem-level shifts despite the lack of exact equivalence.

\begin{table*}[t!]
\footnotesize
\centering
\caption{Bug Category Distribution Across Analyzed TypeScript Projects}
\label{tab:bug_frequencies}
\resizebox{0.85\textwidth}{!}{%
\begin{tabular}{lrrrrrrrrrrrr}
\toprule
\textbf{Project} &
\makecell{\textbf{Async /}\\\textbf{Event}} &
\makecell{\textbf{Error}\\\textbf{Handling}} &
\makecell{\textbf{Missing}\\\textbf{Case}} &
\makecell{\textbf{Missing}\\\textbf{Feature}} &
\makecell{\textbf{Runtime}\\\textbf{Exception}} &
\makecell{\textbf{Tooling /}\\\textbf{Config}} &
\makecell{\textbf{Type}\\\textbf{Error}} &
\makecell{\textbf{Logic}\\\textbf{Error}} &
\makecell{\textbf{API}\\\textbf{Misuse}} &
\makecell{\textbf{Test}\\\textbf{Fault}} &
\makecell{\textbf{UI}\\\textbf{Bug}} &
\textbf{Total} \\
\midrule

refly & 3 & 2 & 3 & 1 & 0 & 1 & 5 & 5 & 11 & 2 & 9 & 42 \\
llamaindexts & 2 & 0 & 4 & 3 & 0 & 16 & 4 & 1 & 3 & 2 & 2 & 37 \\
insomnia & 2 & 1 & 4 & 8 & 0 & 6 & 3 & 2 & 11 & 6 & 2 & 45 \\
xyflow & 1 & 0 & 0 & 0 & 0 & 3 & 7 & 3 & 5 & 0 & 14 & 33 \\
cli & 0 & 0 & 3 & 3 & 2 & 14 & 4 & 9 & 4 & 2 & 1 & 42 \\
hocuspocus & 4 & 0 & 4 & 1 & 0 & 12 & 4 & 0 & 5 & 4 & 3 & 37 \\
n8n & 2 & 1 & 9 & 3 & 0 & 3 & 2 & 11 & 9 & 3 & 3 & 46 \\
xstate & 3 & 0 & 1 & 0 & 0 & 2 & 14 & 5 & 0 & 2 & 4 & 31 \\
shortest & 2 & 1 & 3 & 2 & 0 & 10 & 4 & 1 & 4 & 16 & 2 & 45 \\
user-event & 1 & 0 & 0 & 0 & 0 & 1 & 7 & 4 & 16 & 8 & 7 & 44 \\
type-challenges & 0 & 0 & 10 & 1 & 0 & 1 & 6 & 1 & 2 & 10 & 5 & 36 \\
nest & 0 & 0 & 0 & 0 & 0 & 32 & 1 & 2 & 3 & 1 & 1 & 40 \\
vue-i18n & 1 & 0 & 5 & 2 & 0 & 15 & 5 & 6 & 5 & 1 & 0 & 40 \\
chakra-ui & 0 & 0 & 0 & 0 & 0 & 15 & 5 & 1 & 5 & 0 & 13 & 39 \\
fabricjs & 0 & 0 & 6 & 0 & 2 & 1 & 2 & 8 & 4 & 0 & 10 & 33 \\
shadcn-table & 0 & 0 & 0 & 0 & 0 & 40 & 2 & 0 & 1 & 0 & 0 & 43 \\

\midrule
Mean & 1.31 & 0.31 & 3.25 & 1.5 & 0.25 & 10.75 & 4.69 & 3.69 & 5.5 & 3.56 & 4.75 & 39.56 \\
Total & 21 & 5 & 52 & 24 & 4 & 172 & 75 & 59 & 88 & 57 & 76 & 633 \\ 
 \bottomrule
\end{tabular}
}
\end{table*}

\subsection{Reliability and Validity}
\label{sec:validity}

To ensure labeling consistency, we employed independent double annotation, consensus refinement, and manual verification of model predictions.  
The hybrid labeling and validation pipeline supports reproducibility while retaining human interpretive accuracy.  
A detailed discussion of threats to validity is provided in \autoref{sec:threats}.

\section{RQ1: Categories and Prevalence of Bugs}
\label{sec:rq1results}

\begin{figure*}[t!]
\centering
\includegraphics[width=0.95\textwidth]{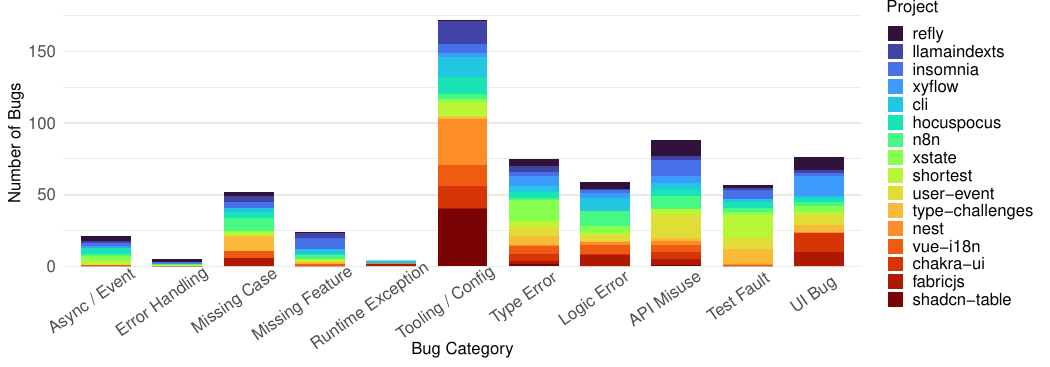}


\includegraphics[width=0.95\textwidth]{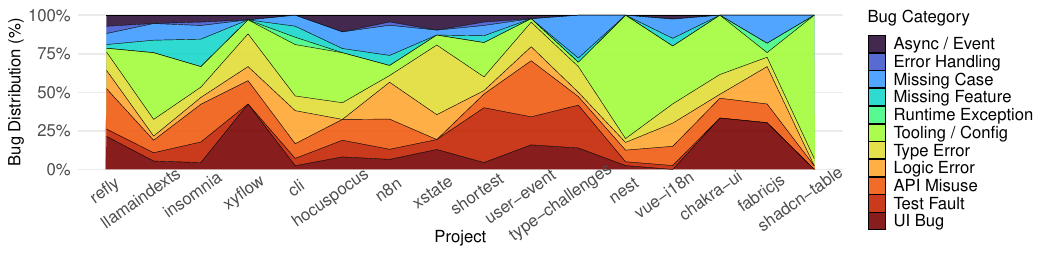}

\caption{Bug category distributions across and within projects. 
\emph{Top:} Overall prevalence of each bug category across all analyzed \ts projects, with stacked segments showing per-project contributions. 
\emph{Bottom:} Normalized distribution of bug categories within each project, showing the relative composition of fault types across systems.}
\label{fig:bug_frequencies_combined}
\end{figure*}

We identified eleven recurring categories of bugs across \numOfProjects{} \ts projects (\numOfBugs total), forming an empirically-grounded taxonomy that captures both \ts-specific and general faults. The taxonomy spans the full ecosystem; from asynchronous and integration faults to type-system, configuration, and logic-level defects. Table~\ref{tab:taxonomy} summarizes the categories with representative examples, while \autoref{fig:bug_frequencies_combined} illustrates their quantitative distribution.

\subsection{Bug Categories}
\label{sec:categories}


\paragraph{Asynchrony / Event.}
Faults stemming from incorrect or incomplete handling of asynchronous operations or events, such as missing \code{await} statements, race conditions, or mismanaged event lifecycles. These errors are often subtle, nondeterministic, and difficult to reproduce. Common patterns include (1) missing or misplaced \code{await} calls leading to premature execution, (2) race conditions or ordering issues in event-driven and asynchronous handlers, and (3) incorrect event registration or propagation causing duplicated or missed events.

\paragraph{Error Handling.}
Failures caused by missing, inadequate, incorrect, or overly-broad exception handling.
They occur when errors are silently ignored or are caught with insufficient recovery logic.
Representative subcategories include missing \code{try/catch} constructs, unhandled promise rejections, 
improper input validations, absence of fallback defaults and recovery mechanisms, and incorrect error propagation or masking during recovery.

\paragraph{Missing Case.}
Incomplete handling of all possible inputs or states, such as missing branches for certain \code{enum} values or neglected edge conditions.

\paragraph{Missing Feature.}
Unimplemented functionality expected by a specification or user interface, such as unbound event handlers or missing API endpoints.

\paragraph{Runtime Exception.}
Defects that trigger crashes or halts during execution, typically due to unsafe state access or violated invariants. These include \code{null}/\code{undefined} dereferences, race-induced runtime failures, infinite loops or deadlocks, and lifecycle violations (e.g., accessing components after disposal).

\paragraph{Tooling / Config.}
Build or compilation misconfigurations that disrupt project behaviour, such as incorrect \code{tsconfig.json} paths, incompatible Babel/Webpack settings, or misconfigured testing environments. These often propagate silently across the toolchain and are difficult to detect through testing.

\paragraph{Type Error.}
Defects caused by incorrect, incomplete, or unsafe type annotations. Common cases include (1) missing or overly-generic types leading to \code{any} inference, (2) inconsistencies between declared and actual types, (3) unsafe casts or misuse of generics, and (4) nullability errors from mismanaged optional types. Such faults highlight the tension between static typing and interoperability with untyped \js code.

\paragraph{Logic Error.}
Traditional control-flow or semantic mistakes unrelated to typing or APIs, such as off-by-one errors, incorrect conditions, or faulty data transformations.

\paragraph{API Misuse.}
Incorrect usage of internal or third-party APIs, including wrong argument order, invalid calls, or misuse of library interfaces (e.g., calling \code{Array.map()} on a string).

\paragraph{Test Fault.}
Bugs within test code or configuration rather than production logic. Subcategories include (1) invalid test setup or configuration (e.g., missing mocks, incorrect imports), (2) flaky or nondeterministic asynchronous tests, (3) incorrect assertions or outdated test data, (4) deprecated and unmaintained tests, and (5) faulty logic in test code. Such issues undermine test reliability and mask real faults.

\paragraph{UI Bug.}
Inconsistent or incorrect user interface behaviour caused by faulty rendering logic, event handling, or state synchronization. Common patterns include visual or layout inconsistencies, incorrect state updates due to stale closures or improper hook usage, and rendering or interaction failures in reactive frameworks.

Overall, this taxonomy reveals a hybrid fault landscape that blends traditional logic and semantic errors with fragilities introduced by \ts's build and type ecosystem. Many bugs emerge at the interfaces between asynchronous behaviour, configuration tooling, and type annotations, highlighting areas where reliability still depends on developer discipline and ecosystem consistency.

\subsection{Bug Prevalence}

\autoref{tab:bug_frequencies} and \autoref{fig:bug_frequencies_combined} summarize the distribution of the \numOfCategories bug categories across all projects, which is highly imbalanced: a few categories account for most observed faults, while others are rare.

\paragraph{Overall trends.}
The most prevalent categories are \emph{Tooling / Config} (27.8\%), \emph{API Misuse} (14.5\%), and \emph{Type Errors} (12.4\%), together representing nearly two-thirds of all bugs. This pattern underscores the fragility of integration boundaries, where build settings, dependencies, and type definitions interact. In particular, the dominance of configuration-related issues reflects the layered and extensible nature of the \ts toolchain, where compiler, transpiler, and bundler configurations jointly determine runtime behaviour. Small inconsistencies in these files can propagate silently through the build pipeline. Such faults are difficult to detect via traditional testing, suggesting the need for more robust CI-based or static configuration checks.

\emph{API Misuse} appears as the second most common category (14.5\%), consistent with previous observations on API fragility in \js and npm ecosystems\cite{hanam2016discovering,bogart2015breaking}. 
The strong reuse culture of npm encourages modular design but also increases dependency coupling, which combined with frequent interface evolution, make such faults difficult to avoid. 
\emph{Type Errors} (12.4\%) remain frequent despite static typing, often caused by unsafe casts, missing annotations, or reliance on \code{any}. These decisions are often made for expediency or compatibility, especially due to \ts's interoperability with untyped \js. This aligns with prior findings on “type erosion” in gradually typed systems~\cite{Bogner_2022, Troppmann_2024}, suggesting that type guarantees are frequently relaxed in practice for convenience or compatibility.

\paragraph{Front-end and semantic bugs.}
\emph{UI Bugs} (12.5\%) are also common, reflecting the complexity of reactive frameworks where component state, Document Object Model (DOM) updates, and asynchronous effects must remain synchronized. Traditional semantic issues such as \emph{Logic Errors} (9.7\%) and \emph{Missing Case / Missing Feature} (12.4\% combined) remain relevant, showing that static typing mitigates, but does not eliminate, high-level reasoning or design faults. These patterns correspond to the “semantic-level” bugs described in prior \js studies~\cite{Ocariza_2013}. 

\paragraph{Low-level and asynchronous faults.}
\emph{Runtime Exceptions} (1.2\%) and \emph{Error Handling} (0.8\%) are comparatively rare, indicating that \ts's typing and defensive idioms (e.g., optional chaining, null guards) effectively prevent many crash-inducing issues. However, \emph{Asynchrony / Event Handling Bugs} (3.6\%) remain nontrivial despite language-level support for \code{async}/\code{await}, confirming that temporal coordination remains a source of subtle errors~\cite{ganji2023jscope}.

\paragraph{Testing-related bugs.}
\emph{Test Faults} (4.5\%) are an emerging class introduced by this study, arising from unstable or incorrect test code. Flaky asynchronous tests and misconfigured mocks are particularly common. These issues reveal that, while testing frameworks such as Jest and Mocha are mature, their interaction with asynchronous workflows remains fragile; an underexplored aspect in prior taxonomies.

\paragraph{Summary.}
Overall, modern \ts projects fail less due to traditional algorithmic mistakes and more due to integration and configuration complexity. The prevalence of build, dependency, and type-level faults indicates that reliability increasingly depends on maintaining consistency across interconnected tools and layers. This shift, from logic-level to ecosystem-level fragility, reflects a broader evolution in web development.
The results motivate our next research questions, which examine how project characteristics and domains influence these patterns (RQ2), and how web application bugs have evolved from \js to \ts (RQ3).

\begin{tcolorbox}[colframe=black!60, title={\textbf{RQ1 Key Findings}}, fonttitle=\bfseries]
Most bugs in modern \ts projects stem from integration and configuration fragility rather than algorithmic mistakes.  
\emph{Tooling / Configuration Issues}, \emph{API Misuse}, and \emph{Type Errors} together account for nearly two-thirds of all defects, reflecting fragility at the boundaries between code, build tools, and external dependencies.  
Traditional errors such as \emph{Logic Errors} and \emph{Runtime Exceptions} are relatively rare, suggesting that  static typing reduces low-level faults, but does not prevent higher-level integration, testing, and asynchronous failures.
\end{tcolorbox}

\section{RQ2: Project- and Ecosystem-Level Correlations}
\label{sec:rq2results}

RQ1 showed that most faults in modern \ts projects arise not from core logic but from
integration and configuration layers.
The prominence of \emph{Tooling / Configuration} and \emph{API Misuse} bugs indicates
that project scale, ecosystem composition, and build infrastructure strongly shape
fault patterns.
Building on this, RQ2 examines which project- and ecosystem-level factors explain
these patterns and how bug categories co-occur across projects.

We analyze relationships between bug categories and intrinsic project attributes
(e.g., size, age, activity, domain) as well as extrinsic ecosystem properties
(e.g., dependency diversity, build tools, and configuration complexity).
Specifically, we (i) compute Spearman correlations between project metrics and
category frequencies, (ii) apply Kruskal--Wallis tests to identify domain-specific
overrepresentations, (iii) analyze cross-category correlations to uncover recurring
\emph{fault clusters}, and (iv) relate dependency characteristics to fault types.

Together, these analyses extend RQ1’s descriptive taxonomy into an explanatory view
of the \ts fault landscape, linking project characteristics and ecosystem complexity
to the emergence of distinct fault patterns.

\subsection{Correlation with Project Attributes (RQ2a.1)}
\label{sec:rq2_attributes}

\begin{figure}[t]
  \centering
  \includegraphics[
    width=0.8\columnwidth,
    trim=0 12 0 12,
    clip
  ]{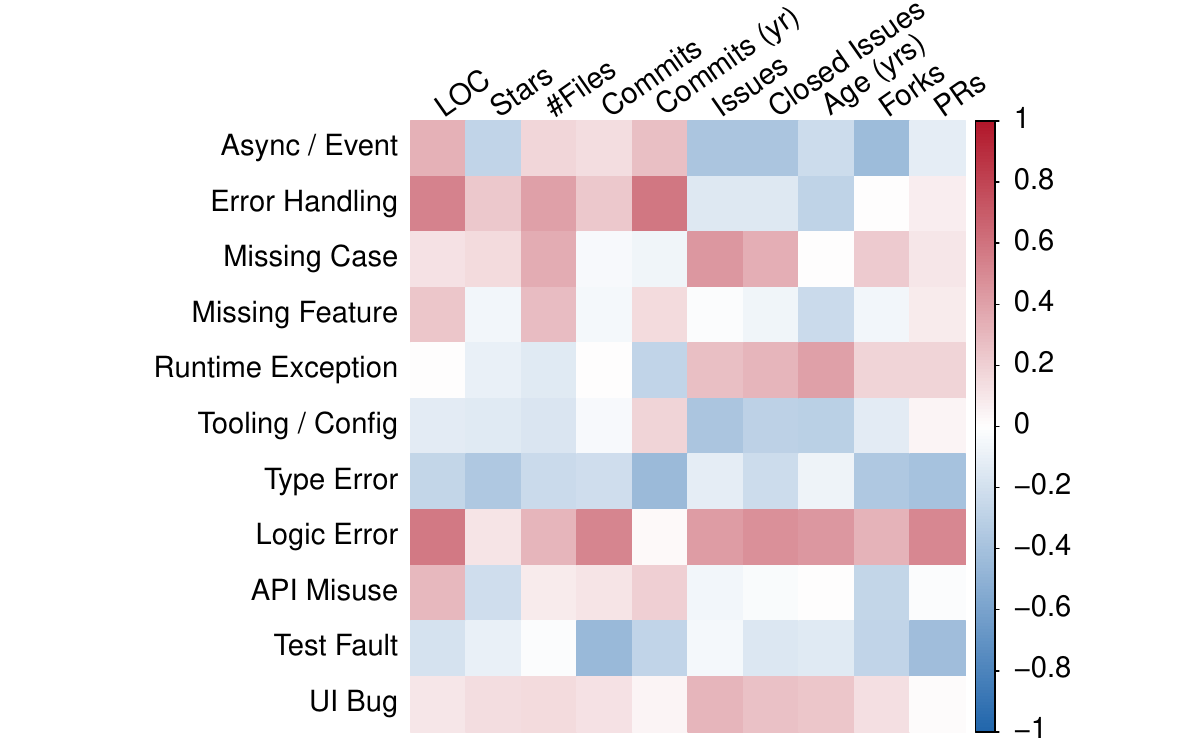}
  \caption{Project attributes and bug categories correlations.}
  \label{fig:bug_project_corr}
\end{figure}

Correlating bug frequencies with project metrics (\autoref{fig:bug_project_corr}) shows that project size, age, and activity exert stronger influence on fault composition than domain alone.  
Several consistent patterns emerge, linking system growth and maturity to distinct categories of faults.

\emph{Error handling, logic, and feature-related bugs} all increase with project size and evolution.  
\emph{Error Handling} correlates with size ($\rho{=}0.52$), commits ($\rho{=}0.49$), and age ($\rho{=}0.44$), indicating that larger, older systems accumulate asynchronous and exception-related issues as recovery logic diverges over time.  
\emph{Logic Errors} also correlate with LOC ($\rho{=}0.49$), consistent with the rise in control-flow complexity and semantic inconsistencies as projects expand.  
Similarly, \emph{Missing Feature} faults correlate with commits ($\rho{=}0.48$) and age ($\rho{=}0.45$), reflecting deferred implementation and evolving requirements typical of long-lived, actively maintained systems.

\emph{Ecosystem and dependency-driven faults} follow a related pattern.  
\emph{API Misuse} correlates with commits ($\rho{=}0.46$), suggesting that frequent updates and dependency churn increase the likelihood of broken API contracts.  
\emph{Tooling / Configuration} issues correlate positively with popularity (stars, $\rho{=}0.42$) and modularity (files, $\rho{=}0.41$), but slightly negatively with LOC ($\rho{\approx}-0.40$).  
These trends imply that configuration fragility grows with ecosystem complexity rather than sheer code volume, as widely-used or modular projects depend on intricate CI/CD pipelines and multi-tool build processes.

In contrast, \emph{Test Faults} and \emph{UI Bugs} exhibit inverse correlations.  
\emph{Test Faults} decrease with project size ($\rho{=}-0.45$), commits ($\rho{=}-0.41$), and popularity ($\rho{=}-0.43$), indicating that mature systems with stable test infrastructure experience fewer test-related failures.  
\emph{UI Bugs} also decline with size ($\rho{=}-0.39$), reflecting that backend- and framework-oriented projects contain fewer UI components and thus fewer opportunities for rendering or state-management failures.

Overall, these correlations reveal a shift in fault composition as projects grow:  
younger or smaller systems exhibit more local logic and testing issues, whereas mature and widely-used projects experience increasing integration and configuration fragility.  
Complexity in modern \ts systems arises less from code volume than from the coordination of asynchronous, tool-driven workflows that link multiple layers of the ecosystem.


\begin{figure}[t]
  \centering
  \includegraphics[
    width=0.9\columnwidth,
    trim=0 8 0 18,
    clip
  ]{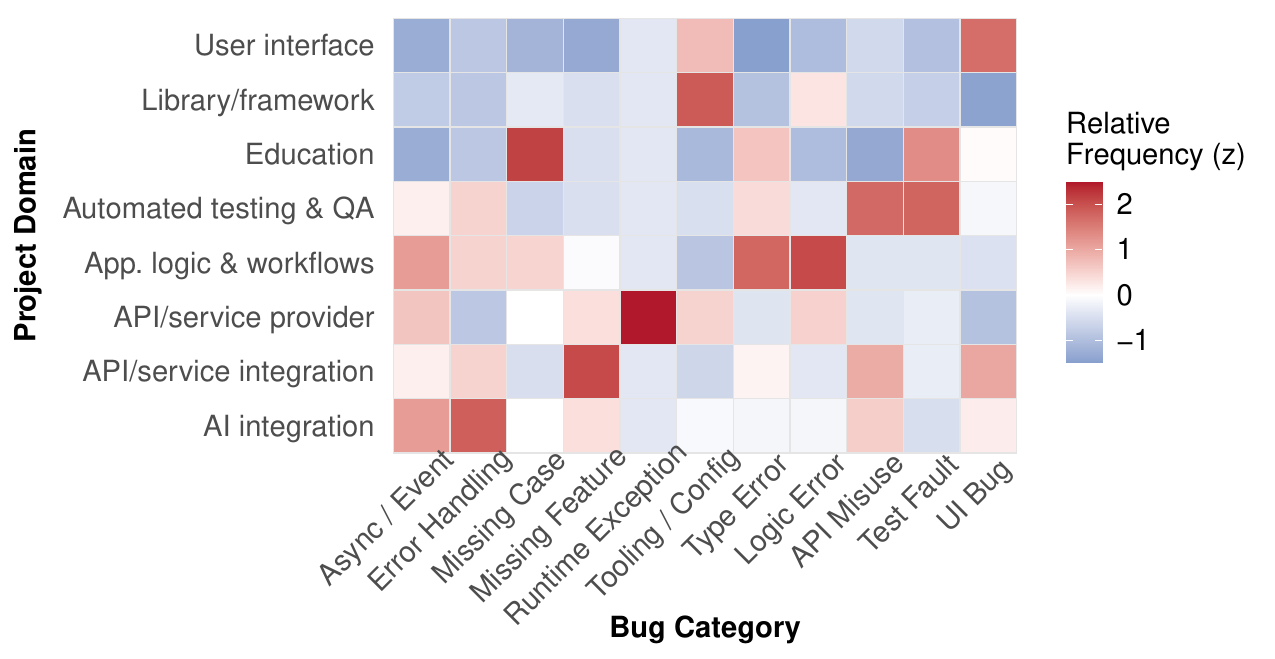}
  \caption{Standardized (z-scored) median frequency of each bug category per domain, showing relative over- or underrepresentation across domains.}
  \label{fig:bug_domain_corr}
\end{figure}

\subsection{Domain-Level Tendencies (RQ2a.2)}
\label{sec:rq2_domain}

We next examined whether bug distributions vary across application domains using Kruskal–Wallis tests followed by z-score normalization of category medians.

\paragraph{Statistical summary.}
No statistically significant differences were detected at $\alpha{=}0.05$, though \emph{Async / Event} ($p{=}0.11$) and \emph{Test Fault} ($p{=}0.10$) showed marginal domain variation, suggesting limited but non-negligible domain influence.  
Future analyses should apply multiple-comparison corrections to mitigate potential false positives and strengthen inference reliability.

\paragraph{Descriptive trends.}
Although statistical significance is limited, the normalized results (\autoref{fig:bug_domain_corr}) reveal several consistent qualitative patterns.  
\textbf{UI-- and frontend-oriented frameworks} (e.g., \textit{chakra-ui}, \textit{fabric.js}) exhibit elevated rates of \emph{UI Bug}, \emph{Async / Event}, and \emph{Type Error} categories, reflecting reactive-state complexity and the tight coupling between event logic and rendering pipelines.  
\textbf{Infrastructure and API-centric systems} (e.g., \textit{nest}, \textit{n8n}) display more \emph{Tooling / Config}, \emph{API Misuse}, and \emph{Runtime Exception} faults, consistent with the demands of dependency integration, build orchestration, and CI/CD configuration.  
\textbf{AI integration} projects (e.g., \textit{LlamaIndexTS}, \textit{refly}) tend to accumulate \emph{Missing Feature} and \emph{Error Handling} bugs, likely due to rapid API evolution and the instability of external model interfaces.  
Finally, \textbf{testing utilities} (e.g., \textit{user-event}, \textit{shortest}) slightly overrepresent \emph{Test Faults}, reflecting continuous adaptation of their test harnesses and high reliance on asynchronous test logic.

Together, these findings suggest that domain alone is not a strong statistical predictor of fault patterns, yet domain-specific architectures influence how faults manifest.  
UI-intensive projects often fail due to runtime synchronization and reactive-state inconsistencies, while infrastructure-heavy systems tend to fail in configuration and integration logic.  
Overall, architectural coupling, rather than domain category itself, emerges as the dominant factor shaping fault tendencies in modern \ts ecosystems.

Taken together, the attribute and domain analyses indicate that project maturity, architecture, and ecosystem complexity collectively drive fault distribution.  
However, these analyses treat each category independently.  
In practice, faults frequently co-occur or cascade. For instance, configuration fragility may amplify type or API misuse errors.  
To capture such interdependencies, the next analysis (RQ2b) examines \emph{cross-category correlations}, identifying recurring \emph{fault clusters} that reflect shared causes or propagation paths within \ts projects.

\subsection{Correlations Among Bug Categories (RQ2b)}
\label{sec:rq2-bug-correlations}

\begin{figure}[t]
  \centering
  \includegraphics[
    width=0.7\columnwidth,
    trim=0 60 0 60,
    clip
  ]{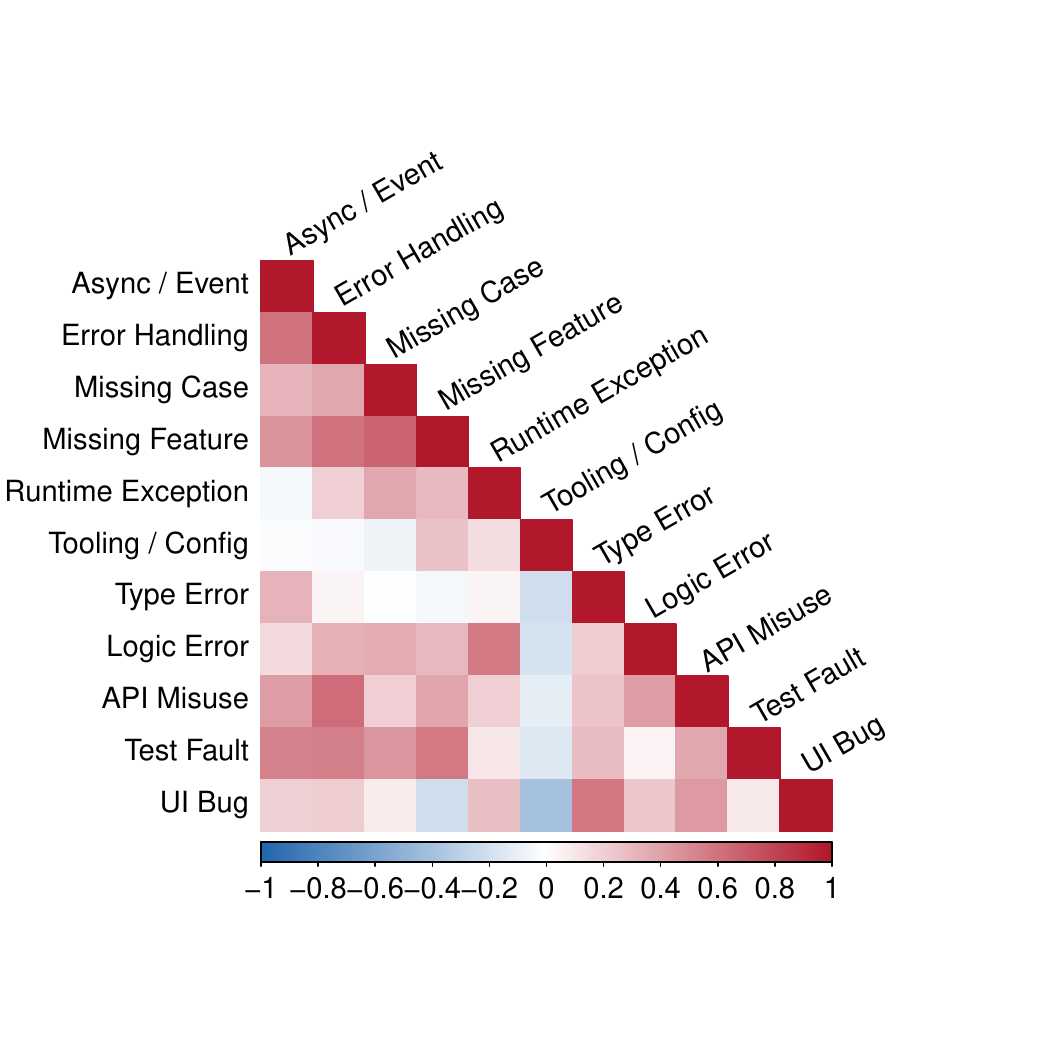}
  \caption{Correlation heatmap among bug categories.
  Colour represents the strength and direction of correlations ($\rho$).}
  \label{fig:bug_bug_corr}
\end{figure}

Pairwise correlations among bug categories (\autoref{fig:bug_bug_corr}) show that fault types in \ts projects are interdependent, forming two main dimensions: an \emph{integration and coordination axis} and an \emph{inverse build--runtime axis}.  
These relationships indicate that many defects arise from the interaction of related failure modes rather than isolated mistakes.

\paragraph{Reinforcing fault patterns.}
Several moderate to strong positive correlations ($\rho>0.4$) reveal reinforcement among conceptually related categories.  
\emph{Missing Feature} correlates with \emph{Missing Case} ($\rho{=}0.62$) and \emph{Error Handling} ($\rho{=}0.51$), showing that incomplete functionality, unhandled states, and weak recovery logic often co-occur.  
\emph{Error Handling} also correlates with \emph{API Misuse} ($\rho{=}0.49$) and \emph{Async / Event} ($\rho{=}0.50$), confirming that asynchronous APIs amplify propagation faults, while \emph{Async / Event} relates to \emph{Logic Error} ($\rho{=}0.43$), indicating the added complexity of asynchronous control flows.  
\emph{API Misuse} correlates with \emph{Type Error} ($\rho{=}0.46$) and \emph{Type Error} with \emph{Logic Error} ($\rho{=}0.42$), linking semantic misuse of libraries with imprecise type inference.  
Finally, \emph{Tooling / Config} correlates with \emph{Runtime Exception} ($\rho{=}0.46$), showing that build misconfigurations can cascade into runtime failures.

\paragraph{Inverse relationships.}
Negative correlations highlight contrasting fault modes.  
\emph{Tooling / Config} inversely correlates with \emph{UI Bug} ($\rho{=}-0.65$) and \emph{Logic Error} ($\rho{=}-0.42$), reflecting a polarity between infrastructure-heavy frameworks and UI-centric applications.  
\emph{Test Faults} also correlate negatively with configuration issues ($\rho{\approx}-0.4$), suggesting that stable CI/CD pipelines reduce test fragility.  
A weaker inverse relation between \emph{Runtime Exception} and \emph{Type Error} ($\rho{=}-0.32$) implies that static typing mitigates, but does not eliminate, some runtime crashes.

\paragraph{Emergent clusters.}
Collectively, the correlations form two coherent clusters.  
The \textbf{integration and coordination cluster} (including \emph{Async / Event}, \emph{Error Handling}, \emph{Missing Case / Feature}, and \emph{Runtime Exception} ($\rho{\approx}0.4$--$0.7$)) captures asynchronous and multi-module fragility, where incomplete implementations and weak recovery logic combine to cause cascading faults.  
The \textbf{inverse build--runtime cluster}, linking \emph{Tooling / Config} and \emph{Type Error} (positively) but inversely related to \emph{UI Bug} and \emph{Test Fault}, reflects a trade-off between build-time robustness and runtime reliability.  
This polarity mirrors the ecosystem’s structural divide: libraries and frameworks experience configuration and typing issues, while user-facing applications exhibit behavioural and runtime faults.

\paragraph{Interpretation.}
Overall, \ts fault patterns organize along two orthogonal dimensions:  
(1) an \textbf{integration axis}, driven by asynchronous coordination, incomplete recovery, and API coupling; and  
(2) a \textbf{toolchain axis}, defined by configuration and build fragility.  
Static typing reduces low-level crashes but shifts fragility toward ecosystem boundaries, where reliability depends on consistent tooling and disciplined API use.  
These relationships suggest that unreliability in modern \ts projects stems from shared mechanisms across components and layers, particularly at the boundaries between asynchronous control, dependencies, and build infrastructure.

To further explore these dependencies, RQ2c analyzes how external ecosystem factors (such as build tooling, dependency composition, and configuration diversity) correlate with specific fault types, extending our view from intra-project dynamics to broader infrastructural influences.

\subsection{Dependency and Toolchain Correlations (RQ2c)}
\label{sec:rq2-dependency-correlations}

\begin{figure}[t]
  \centering
  \includegraphics[
    width=0.85\columnwidth,
    trim=0 8 0 0,
    clip
  ]{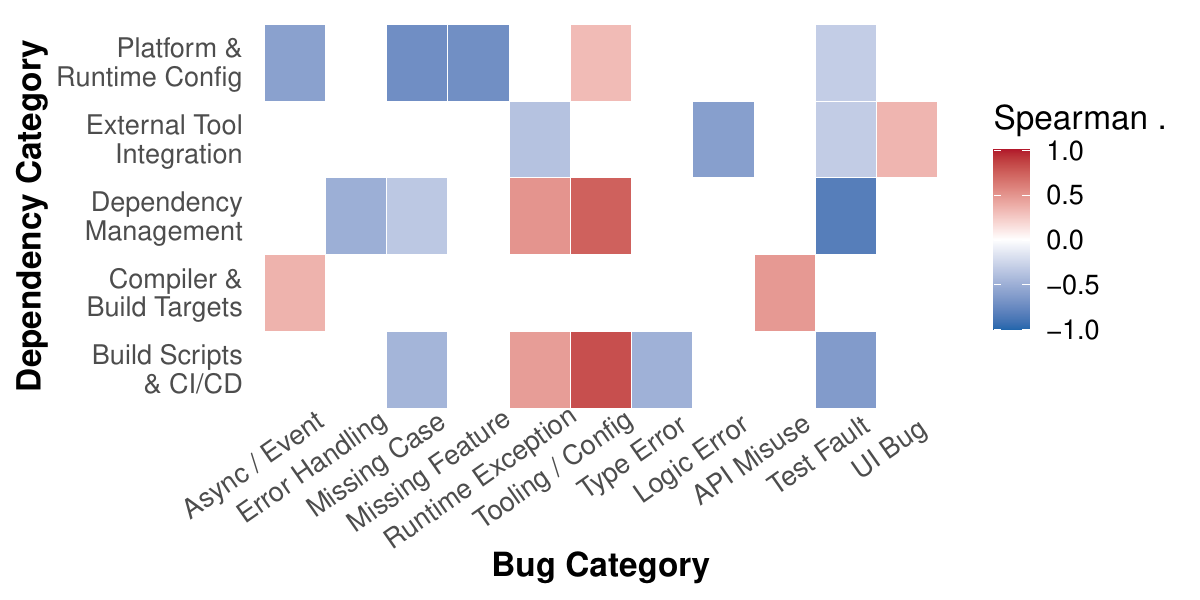}
  \caption{Correlations between dependency categories and bug types.
  Colour shows correlation strength and direction.}
  \label{fig:dependency_bug_corr}
\end{figure}

The analysis of dependency and toolchain composition shows that configuration fragility in \ts projects stems primarily from the \emph{depth of build automation} and \emph{heterogeneity of toolchains}, rather than code size or domain complexity.

\paragraph{Category-level correlations.}
\autoref{fig:dependency_bug_corr} shows Spearman correlations between dependency categories and bug types.  
The strongest positive associations appear in the \emph{Tooling / Config} dimension, confirming that configuration faults arise from layered automation and diverse toolchains.  
\emph{Build Scripts \& CI/CD} correlate most strongly with \emph{Tooling / Config} bugs ($\rho{=}0.77$), indicating that chaining multiple linters, formatters, or release scripts introduces integration errors across build stages.  
\emph{Compiler \& Build Targets} ($\rho{=}0.66$) correlate with both \emph{Tooling / Config} and \emph{Type Error} categories, reflecting the challenge of maintaining consistent transpiler configurations (e.g., \ts, Babel, Rollup).  
\emph{Platform \& Runtime Configuration} ($\rho{=}0.61$) associate with \emph{Runtime Exception} and \emph{Error Handling} bugs, showing that environment-specific settings (browser, Node.js, serverless) often fail at execution time.  
\emph{External Tool Integration} ($\rho{=}0.46$) moderately correlates with configuration and API-related issues, while \emph{Dependency \& Package Management} exhibits only weak correlations, suggesting that dependency count alone is a poor predictor of fault proneness compared to build complexity.

\paragraph{Individual dependencies.}
At the package level, developer tools such as \emph{eslint}, \emph{prettier}, and \emph{commitlint} correlate moderately with \emph{Tooling / Config} bugs ($\rho{=}0.43$--$0.54$), confirming that integrating multiple automation tools increases CI/CD complexity.  
Similarly, orchestration frameworks (\emph{babel}, \emph{rollup}, \emph{jest}) show positive associations with configuration and type-related faults, reflecting the tension between flexibility and consistency in the \ts build ecosystem.

\paragraph{Aggregated trends.}
Across dependency categories, configuration fragility scales not with dependency count but with the \emph{diversity of roles} and \emph{depth of automation}.  
Projects combining multiple compilers, linters, and test harnesses exhibit ``toolchain entanglement,'' where a misconfiguration in one layer propagates downstream.  
Conversely, libraries with simpler or fewer build stages show lower configuration fault rates even when large.

\paragraph{Interpretation.}
These results reflect a broader shift in \ts development: major faults now stem from \textbf{ecosystem-level coordination failures} rather than code-level complexity.  
Build, packaging, and automation systems form dense dependency networks where minor inconsistencies cascade through the pipeline.  
Improving reliability therefore requires reducing configuration coupling (via unified build metadata, automated consistency checks, and compatibility validation) rather than focusing solely on type precision or static analysis.

\subsection{Synthesis and Implications}
\label{sec:rq2-synthesis}

Across all analyses, RQ2 shows that the fault landscape of \ts projects is driven less by code-level complexity than by ecosystem coordination.  
Most recurring failures arise at the boundaries between components, build pipelines, and dependencies rather than within local program logic.  
Integration, configuration, and asynchrony remain persistent fault sources despite strong static typing guarantees: correctness in modern \ts systems depends on consistent alignment among compilers, build tools, and runtime frameworks.

While static typing reduces traditional runtime and logic faults, it displaces failure risk upward into orchestration layers.  
Misconfigurations, version drift, and asynchronous timing errors persist because static guarantees cannot enforce cross-tool consistency.  
This shift from code-level to ecosystem-level fragility marks a structural transformation in how reliability risks manifest in typed web systems.

Architectural and evolutionary effects reinforce this pattern.  
Larger and older projects show more \emph{Error Handling} and \emph{Missing Feature} bugs (evidence of accumulated technical debt) whereas mature systems with stable CI/CD pipelines report fewer \emph{Test Faults}.  
Frameworks and infrastructure libraries are dominated by build-time and configuration issues, while UI-centric applications experience more runtime and asynchronous failures.  
This polarity reflects the specialization of the \ts ecosystem, where each abstraction layer introduces distinct failure modes.

Dependency analyses confirm that fault risk scales with automation depth and heterogeneity rather than dependency count.  
Projects combining multiple compilers, linters, and test frameworks exhibit ``toolchain entanglement,'' where errors in one layer propagate across the pipeline.  
Simpler build setups remain more robust even when large, underscoring the need to treat configuration and orchestration as first-class reliability concerns.

Overall, \ts shifts the locus of failure from logic to orchestration: static types prevent many local errors but amplify dependence on coherent toolchains and consistent configuration across compilers and frameworks.  
Improving reliability therefore requires ecosystem-level strategies: automated detection of configuration drift and metadata misalignment, integration of asynchronous and runtime analyses with static checks, and systematic validation of CI/CD consistency.  
Reliability in modern web ecosystems thus depends as much on the cohesion of the toolchain as on the correctness of the code itself.

\begin{tcolorbox}[colframe=black!60, title={\textbf{RQ2 Key Findings}}, fonttitle=\bfseries]
\textbf{Ecosystem complexity dominates.} Bug patterns are shaped more by architectural coupling and build orchestration than by project size or domain.\\[2pt]
\textbf{Two principal fault axes.} (1) An \emph{integration axis} linking asynchronous, API, and error-handling faults; and (2) a \emph{toolchain axis} connecting configuration and typing issues.\\[2pt]
\textbf{Shifted fault boundary.} Static typing suppresses logic errors but relocates failure risk to CI/CD pipelines and dependency integration, emphasizing orchestration over algorithmic correctness.
\end{tcolorbox}

The combined results from RQ1 and RQ2 reveal a shift in modern \ts fault characteristics; \textit{from logic and runtime errors to configuration and integration fragility}.
To determine whether this represents a broader evolution of the web ecosystem or a language-specific trend, we next compare \ts faults with those historically observed in JavaScript to assess how static typing and ecosystem maturity have transformed modern web reliability.


\section{RQ3: Evolution of Fault Types and Ecosystem Shifts}
\label{Sec:rq3results}

Building on RQ1 and RQ2, this section examines whether findings from prior studies of \js bugs remain applicable to modern \ts development.
We compare the prevalence of bug categories identified in RQ1 with those reported by \citet{Ocariza_2013}, one of the most widely-cited and representative empirical studies of \js faults.
Our taxonomy is informed by the schemes of \citet{Ocariza_2013} and includes several conceptually-overlapping categories, enabling partial comparability across studies.
Using the category mapping and maximum corresponding prevalence described in \autoref{sec:historical_mapping}, we identify both continuities and systematic shifts in the web application fault landscape.

\paragraph{Common categories.}
Across both languages, bugs related to API misuse, logic, and UI behaviour appear common,
though their underlying causes have evolved.
\citet{Ocariza_2017} found that most \js bugs arose from
incorrect method parameters, 91\% of which were DOM-related~\cite{Ocariza_2017}.
In contrast, UI-related bugs constitute only 12.5\% of our \ts dataset,
suggesting that while similar surface categories persist, the specific sources of error,
particularly those tied to direct DOM manipulation, have diminished.

\paragraph{Decreased prevalence in \ts.}
The most substantial reductions appear in UI-- and type-related faults.  
DOM interaction issues, which accounted for 68\% of client-side JavaScript bugs~\cite{Ocariza_2017}, have largely receded with the adoption of frameworks that abstract DOM management (e.g., React, Angular).  
Similarly, type-related errors fell from $\sim$33\% in \js to 12.4\% in our \ts sample, reflecting the preventive effect of gradual typing and compiler checks.

\paragraph{New and rising categories.}
Conversely, several categories have emerged or grown markedly since earlier \js studies.
\emph{Tooling / Configuration} bugs, the most prevalent in our dataset, and \emph{Test Faults}
have no direct analogues in prior taxonomies.  
Their appearance underscores how modern development practices have shifted from ad-hoc
scripting to highly-automated pipelines.  
These categories reflect new sources of fragility introduced by build systems, dependency
management, and continuous integration tooling.

\subsection{Synthesis and Implications}

The observed changes highlight a structural evolution in web development, rather than a linguistic one.  
The reduction of DOM and type faults corresponds to stronger language guarantees
and widespread use of framework-managed UIs, which mediate most DOM
interactions.
Industry surveys report React alone being used by 80\% of \js developers
and 39.5\% of all developers~\cite{stateofjs2024,stackoverflow2024}.

The emergence of testing and tooling issues likely reflects maturation:
more projects now employ formal testing, but these additional layers introduce new fault opportunities.  
\citet{fard2017javascript} previously found that 40\% of client-side JavaScript projects
lacked tests; by contrast, most modern \ts repositories integrate
Jest, Mocha, or similar frameworks.  
Thus, the presence of test-related defects may signify more widespread testing
rather than poorer quality control.

The dominance of \emph{Tooling / Configuration} faults indicates a major shift in
developer workflow.  
Modern web projects rely on complex ecosystems of compilers, transpilers, and package
managers that extend beyond the code~\cite{jetbrains2024}.
While these tools improve scalability and maintainability, they also introduce new failure modes.
(e.g., misaligned versions, incompatible plugins, or inconsistent build metadata) now reflected
as a leading fault category in our findings.

Overall, these patterns suggest that reliability challenges in \ts no longer stem primarily
from logic or syntax but from coordination across abstraction layers.  
This shift aligns with the broader trend observed in RQ2: software correctness
increasingly depends on the cohesion of the ecosystem rather than the precision
of individual statements.  
The decline in traditional runtime and type faults, alongside the rise of integration
and tooling issues, reflects a maturing yet more interconnected web development environment.

\paragraph{To Type or Not to Type.}
Studies on the impact of types in JavaScript and \ts have yielded mixed results.
\citet{Gao_2017} found that around 15\% of JavaScript bugs could be prevented
with static typing,
while \citet{Bogner_2022} found that although \ts improved
code quality and readability, it did \emph{not}
significantly reduce bug proneness.
Our results offer a possible explanation:
type systems may mitigate traditional logic and runtime errors, but much of today’s
fault density lies outside the scope of typing, and within configuration, dependency,
and testing layers. This displacement helps reconcile why stronger type systems
can enhance code quality without necessarily reducing overall defect counts.

\section{Threats to Validity}
\label{sec:threats}

\textbf{Internal and construct validity:}
Bias may arise from manual labeling and imperfect linkage between issues, commits, and pull requests.
We mitigated this using explicit mining heuristics, double annotation with consensus resolution, and spot-checking of classifier outputs.
The taxonomy captures observed \emph{fault manifestations} rather than root causes; potential overlap between categories (e.g., \emph{Async / Event} and \emph{Error Handling}) was minimized through iterative refinement.
We do not use issue or PR resolution time as a proxy for bug difficulty, as time-to-close is confounded by prioritization and contributor availability.
Project metadata from GitHub may contain missing or inconsistent values.

\textbf{External validity:}
Our dataset includes \numOfProjects diverse, actively-maintained open-source projects; results may not generalize to smaller, inactive, or proprietary systems.
Sampling prioritized domain diversity over exhaustive coverage.

\textbf{Conclusion validity:}
We used non-parametric statistics (Spearman’s~$\rho$, Kruskal--Wallis) suitable for skewed data.
All reported relationships are associative rather than causal, and we report only moderate or stronger effects ($|\rho|\!\ge\!0.4$).
The RQ3 comparison with \js is indirect: prior studies report aggregate results without project-age metadata, preventing one-to-one matching and control for maturity effects.

All scripts, datasets, and analysis artifacts used in this study are publicly available to support reproducibility.\footnote{\url{https://github.com/seatsfu/tsbugs}}
\section{Related Work}
\label{Sec:related-work}

\noindent
\textbf{Historical Studies of Bugs.}
Many prior studies have built or analyzed defect datasets
for languages other than JavaScript or TypeScript~\cite{Li_2006, Just_2014, Tan_2014, Catolino_2019},
They  informed best practices in dataset design~\cite{Bird_2009} and explored how defects arise in tests and infrastructure~\cite{vahabzadeh2015empirical}.
These works highlight challenges in dataset construction and validity,
while our focus is on defects specific to JavaScript, TypeScript, and similar dynamic languages.

\noindent
\textbf{JavaScript Bugs.}
Past work on JavaScript bugs is most closely related to ours.
\citet{Ocariza_2013} examined client-side \js,
and found that most faults stem from DOM interactions~\cite{Ocariza_2011, Ocariza_2013, Ocariza_2017}.
BugsJS focused on 
\emph{server-side} bugs~\cite{Gyimesi_2020}.
\citet{Wang_2017} studied concurrency bugs in Node.js, finding that two-thirds involved atomicity violations.
\citet{selakovic2016performance} identified common causes of performance bugs~\cite{selakovic2016performance},
while \citet{Hashemi_2022} and \citet{Pei_2025} analyzed flaky tests,
linking them to concurrency, asynchrony, and DOM events. 
Like our work, these studies taxonomize and analyze bug patterns,
but focus on JavaScript rather than TypeScript.
Given the language and ecosystem differences, their findings do not necessarily generalize to \ts.

\noindent
\textbf{Types and Dynamic Languages.}
Our study explores how type systems influence faults in typescript
and how these differ from prior JavaScript results.
\citet{Gao_2017} estimated that about 15\% of JavaScript bugs could be avoided
 using TypeScript's type-checking, while
\citet{Bogner_2022} found that TypeScript code exhibited higher quality metrics
but similar bug rates.
\citet{Fischer_2015} reported
productivity gains,
and
\citet{Wilson_2018} observed that developers often find \ts's type erasure semantics unexpected.
These suggest that the JavaScript findings
may not directly transfer to TypeScript,
motivating our investigation.

Other work  compared declared and inferred types in TypeScript 
to detect mismatches~\cite{Feldthaus_2014,Williams_2017},
though these focused on tool development rather than bug characterization.
Parallel research in gradually-typed languages such as Python  examined developer adoption~\cite{Chen_2020}, typing-related defects~\cite{Khan_2021}, and checker effectiveness~\cite{Xu_2023},
offering context for the transition to gradual typing but not replacing the need for dedicated studies of \ts.

\noindent
\textbf{Empirical Studies of JavaScript.}
Other empirical research have examined broader reliability issues in \js,
including code smells~\cite{saboury2017empirical,johannes2019large},
syntactic bug patterns~\cite{hanam2016discovering}, and
testing practices~\cite{fard2017javascript}.
Further studies explored dynamic behaviours such as
type coercions~\cite{pradel2015good},
object usage~\cite{wei2016empirical},
\code{eval}~\cite{Richards_2011_eval},
insecure behaviours~\cite{Yue_2009},
information flows~\cite{staicu2019empirical},
or other dynamic behaviour~\cite{Richards_2010}.
\citet{wang2023empirical} examined bugs in JavaScript engines rather than applications. 

These directions deepen understanding of \js and its limitations relative to \ts, but revisiting them within the \ts ecosystem is beyond the scope of this work.





%

\section{Concluding Remarks}
\label{Sec:conclusion}

This study characterizes fault patterns in modern \ts projects, revealing a shift from
code-level defects toward ecosystem-level fragility.
Across systems, reliability increasingly depends on toolchains,
dependencies, and asynchronous workflows rather than local logic alone.
\emph{Tooling / Configuration}, \emph{API Misuse}, and \emph{Type Errors} dominate, with
fault distributions varying by project maturity and ecosystem complexity.
Compared to prior \js studies, traditional DOM and runtime faults have declined,
while configuration and testing failures now prevail.
Overall, \ts improves local correctness but shifts reliability risks to integration and
orchestration, indicating that modern reliability tools must treat configuration,
dependencies, and asynchrony as first-class concerns.

\section{Data Availability}

The data, scripts, and experimental artifacts used for data collection, processing, and analysis in this study are available  \ulhref{https://github.com/SEatSFU/tsbugs}{here}.
\begin{acks}
This work was supported in part by the Natural Sciences and Engineering Research Council of Canada (NSERC).
We thank Shivang Jain for his assistance with manual bug labeling and validation during taxonomy development.
We also thank the anonymous reviewers for their thoughtful and constructive feedback.
\end{acks}


\balance
\bibliographystyle{ACM-Reference-Format}
\bibliography{refs}


\end{document}